# Thermal stability of a free nanotube from single-layer black phosphorus

Kun Cai [1, 2 *], Jing Wan [1], Ning Wei [1*], Haifang Cai [1], Qing-Hua Qin [2]

[1] *College of Water Resources and Architectural Engineering, Northwest A&F University, Yangling 712100, China*
[2] *Research School of Engineering, the Australian National University, Canberra, ACT, 2601, Australia*
* Corresponding author's email address: kuncai99@163.com (Kun Cai); nwei@nwsuaf.edu.cn (Ning Wei)

**Abstract**:
Similar to the carbon nanotube fabricated from graphene sheet, a black phosphorus nanotube (BPNT) also can theoretically be produced by curling the rectangular single-layer black phosphorus (SLBP). In present study, the effect of thermal vibration of atoms on the failure of a BPNT is investigated using molecular dynamics simulations. Two types of double-shell BPNTs, which are obtained by curling the rectangular SLBP along its armchair/pucker direction and zigzag direction (in-plane normal) respectively, are involved in simulation. At finite temperature, a bond on the outer shell of tube is under tension due to both of curvature of tube and serious thermal vibration of atoms. As the length of a bond with such elongation approaches its critical value, i.e., 0.279 nm, or the smallest distance between two nonbonding phosphorus atoms is over 0.389nm caused by great variation of bond angle, the tube fails quickly. The critical stable states of either an armchair or a zigzag BPNT at finite temperature are calculated and compared. To achieve a stable BPNT with high robustness, the curvature of the tube should be reduced or the tube should work at a lower temperature. Only when the BPNT has structural stability, it has a potential application as a nanowire in a future nano electro-mechanical system (NEMS).

## 1. Introduction

Phosphorus is the No. 15 element in group 5A in the periodic table. Black phosphorus (BP), which is the most stable structure among the phosphorus allotropes [1-4], is a semiconductor material formed with $sp^3$ hybridized bonds. During the past few years [5-11], the single layer of black phosphorous (SLBP), as a new two-dimensional material, attracts much attention due to its unique electronic properties, e.g., over $10^4$ of drain current modulation, modified direct gap, high charge-carrier mobility (up to $10^3$ cm2/(V.s)), etc.

When used as a component in a nano device, the structural configuration of a SLBP should be stable in order for the mechanical stability of the device, which has invoked the understanding of the mechanical properties of SLBP. Using dispersion-corrected density functional theory, Appalakondaiah et al. [12] investigated the van der Waals (vdW) interactions among the atoms in the orthorhombic BP, and presented elastic properties of multi-layer BP. Jiang and Park [13] studied the mechanical parameters of SLBP under uniaxial tension by ab initio calculations. Their results showed that the SLBP is an orthotropic material during elastic deformation. The elastic modulus along the pucker direction, ~106.4GPa, is about 2.5 times of that along the (in-plane) vertical direction. Due to the fact that a $3sp^3$ bonding strength is weaker than that of a $2sp^2$, the modulus of SLBP is far less than that of the single layer graphene (~1Tpa). In particular, Jiang and Park [8] found that the Poisson's ratio of SLBP in the out-of-plane direction was negative when a SLBP is under uniaxial tension along pucker direction. Also using first principles calculations, Hu *et al.* [7] investigated the mechanical and electronic properties of both the SLBP and bilayer BP under either isotropic or uniaxial strain, and found the band gap varies with the deformation. Wei and Peng [14] studied the ideal tensile strength and critical strain of SLBP and proposed a general formulation to calculate the Young's moduli for such 2D structure based on continuum theory. Kou *et al.* [15] studied the ripple deformation in a SLBP induced by compression along pucker direction. Using density functional tight binding calculations, Sorkin and Zhang [16] studied the edge stress, moduli of phosphorene and examined the structure and stability of the phosphorene edges with such stress. Recently, Jiang [17] set up a set of parameters for the Stillinger-Weber (SW) potential from the valence force field model. The accuracy of using these parameters was checked by comparing the phonon spectrum obtained from *ab initio* calculations. In his work, the stress-strain curves of a SLBP under uniaxial tension were given at different temperature. Sha et al. [18] investigated the moduli and fracture limits of SLBP under either uniaxial or bi-axial loading using the SW potential. Yang *et al.* [19] researched the mechanical behavior of SLBP under shear loading at finite temperature.

In the above studies, the nano tube from SLBP is not considered. As a matter of fact, a carbon nanotube can be obtained by curling a sheet of graphene. Similarly, it can be imagined that a BP nanotube (BPNT) might also be obtained by curling a SLBP. Up to now, only a few work has introduced BPNT. For finding new nano structured materials with new properties, Seifert and Hernández [20] once gave a discussion on fabrication of a stable nanotube from such element as phosphorus rather than carbon. In their work, a 2D honeycomb was firstly formed from a BP sheet. Then, a 2D honeycomb sheet was obtained by removing atoms from BP sheet. Final, a tube was fabricated by curling the 2D honeycomb sheet. Guo *et al.* [21] reported the electronic properties of phosphorene based nanostructures, including BPNTs. In their report, two types of BPNTs (Figure 1e and f) were involved. Obviously, mechanical deformation of few-layer BP is significant for its electronic properties. However, the strength of $3sp^3$ bond in SLBP is much lower than that between the $2sp^2$ carbon atoms in graphene. When the radius of a BPNT is too small, the P-P bonds on the outer shell of the tube are easily broken. Especially, at finite temperature, thermal vibration of atoms on a BPNT will reduce the strength of the tube further. Without the stability of BPNT, the other physical properties of such nano structures are meaningless. When curling a SLBP to form a BPNT, both of curling direction and the radius of the curved neutral layer determines its geometric configurations. In this paper, we will present the study of stability of BPNT with respect of critical curvature of the BPNTs shown in Figure 1e and f, and the minimum number of periodic cells along the circular direction of a BPNT at finite temperatures.

## 2. Computational models

Figure 1 gives two types of BPNTs by curling a rectangular SLBP along different directions. For an armchair BPNT (A-BPNT), the whole tube is adopted in simulation. The system is kept at a canonical NVT ensemble and the temperature increases from 4.3 K to specified temperature, e.g., 150 or 300 K, after 1 million time steps, after arriving at the temperature, 10 million more time steps of relaxation is carried out. For a zigzag BPNT, we only choose an arch segment from the tube for simulation. The segment will be relaxed for 10 million time steps at a NVT ensemble with a constant temperature. During molecular dynamics (MD) simulation, near each straight side, only one layer of top atoms and one layer of adjacent bottom atoms are fixed.

To have perfect computation efficiency with reasonable accuracy for nonlinear behavior of a BP nanoscale structure, the SW potential developed by Jiang [17] is chosen to describe the interaction among atoms in the structure [22]. In the present study, we choose LAMMPS [23] for the following simulations. In the time integration, time step is 0.1 fs.

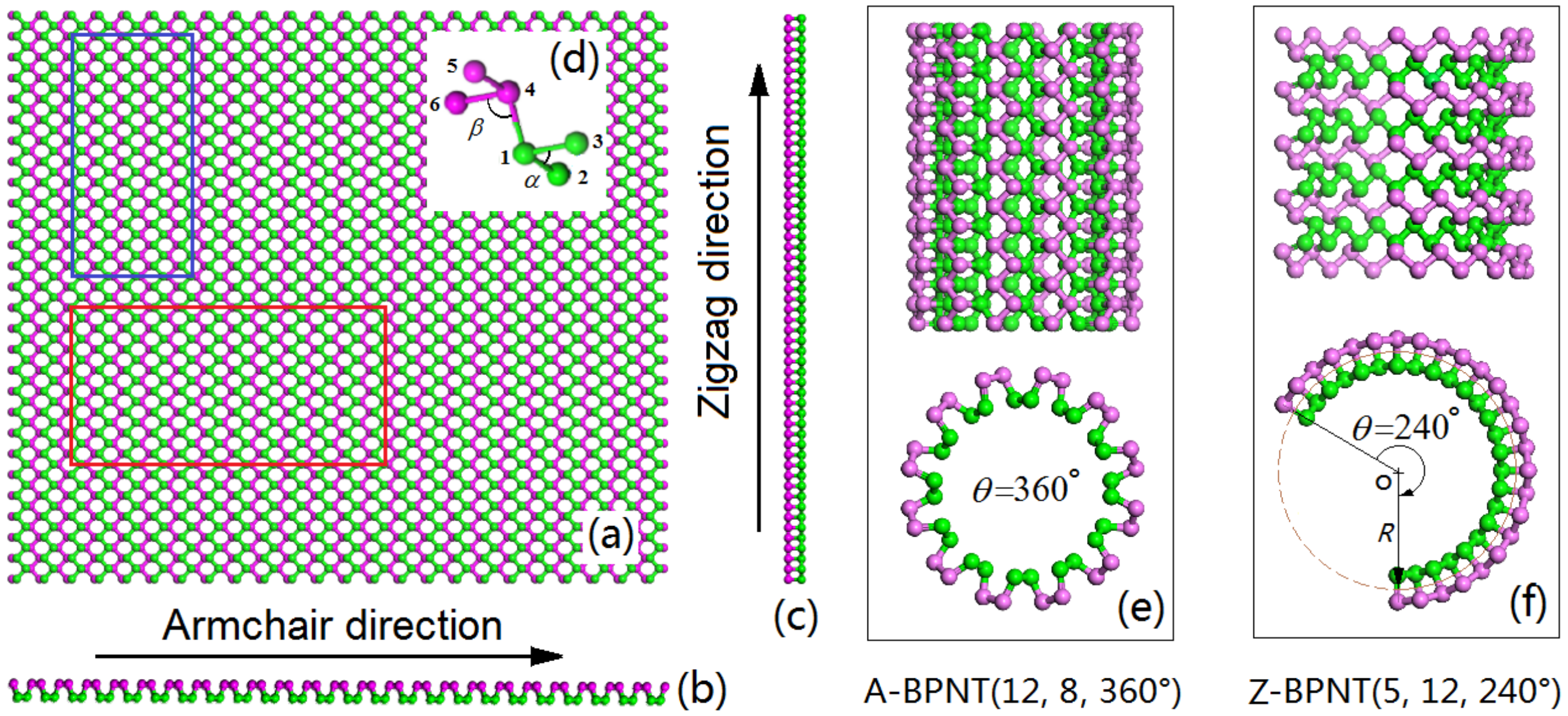

**Figure 1.** Geometric configurations of nano shells from SLBP: (a) top view of SLBP; (b) zigzag-view of SLBP; (c) armchair-view of SLBP; (d) unit cell with bond lengths of $L_{12}$=$L_{13}$= $L_{45}$= $L_{46}$=0.2224 nm; $L_{14}$=0.2244 nm, bond angles of $\alpha = 96.36°$ and $\beta = 102.09°$; (e) Armchair BPNT(A-BPNT), which is formed by curling the armchair sides of SLBP within the red rectangular in (a); (f) Zigzag BPNT(Z-BPNT), which is formed by curling the zigzag sides of SLBP within the blue rectangle in (a). R is the radius of the curved neutral layer (dash curve in (f)). For simplicity, using ($N_A$, $N_Z$, $\theta$) labels the chirality of a BPNT. Where $N_A$ and $N_Z$ are the numbers of periodic cells along the armchair and zigzag directions within the rectangle, respectively. $\theta$ is the curved angle. For example, the A-BPNT in (e) is represented as ($N_A$, $N_Z$, $\theta$) = (12, 8, 360°), and the Z-BPNT in (f) is of (12, 8, 240°). $R$ is the radius of the curved neutral layer (dash line in (f)) of curved shell.

The two models of BPNTs shown in Figure.1 are obtained by ideal geometrical mapping of the SLBP. During mapping, the inner layer (colored green) becomes shorter as compared with its original length, and the outer layer (colored purple) becomes longer. Hence, the bonds on the outer layer are under tension. If the tensile deformation of the outer layer is too higher, the P-P bonds break. Besides, in a practical application, the nano structure should work at finite environmental temperature. Hence, the tube becomes more fragile when thermal vibration of the atoms on the tube is involved. To find a stable BPNT, in the present study, a BPNT obtained by geometrical mapping of SLBP is relaxed in thermal environments. When the tube keeps undamaged during relaxation, it is stable. When the number of periodic cells along circular direction of the tube reaches minimum, the critical radius of the tube is obtained.

# 3 Results and discussions

## 3.1 Critical curvature of a BPNT at different temperature

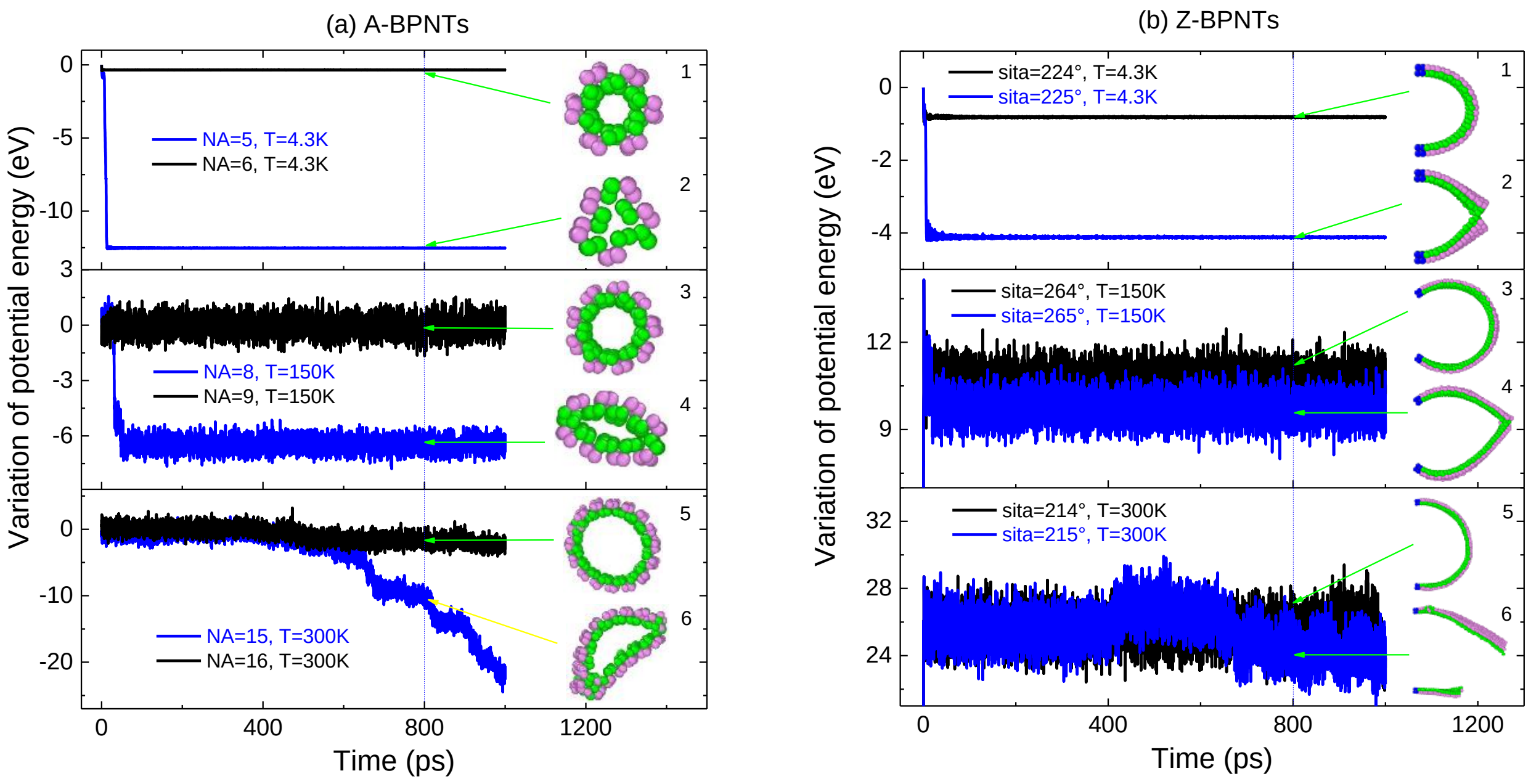

**Figure 2.** Variation histories of potential energy of curved SLBPs along different direction at different environmental temperatures, e.g, 4.3, 150 and 300K: (a) along armchair direction, the A-BPNTs are formed with chirality of ($N_A$, 7, 360°); (b) along zigzag direction, the Z-BPNTs are formed with partial chirality of (6, 15, $\theta$/sita) at 4.3K or (6, 31, $\theta$) at 150 and 300K. The snapshots of related (axial view) models at 800 ps are inserted.

After a series of tests, we find that the minimum number of periodic cells along the circular direction of an A-BPNT is only 6 at 4.3 K (see the top layer in Figure 1a). The minimum number increases to 9 at 150 K or 16 at 300 K. When the number of periodic cells is less than the minimum number, the tube collapses within few ps. For example, at 4.3 K, the A-BPNT with chirality of (**6**, 7, 360°) is stable, however, the A-BPNT (**5**, 7, 360°) becomes a triangular prism (see the second inserted snapshot in Figure 1a). Why does it happen? It is known that the inner layer of the tube becomes shorter after geometrical mapping, and the distances between the atom on the inner layer and its neighbor atoms is reduced. The shorter distances between the non-bonded atoms on the inner layer result in the stronger attraction (vdW interaction), simultaneously, the P-P bonds on the outer tube are under tension due to the curving of the tube. The tube fails to keep a stable configuration under such conditions. As there are only 6 periodic cells along the circular direction of the A-BPNT (**6**, 7, 360°) at 4.3 K, the critical value of the curved angle per periodic cell (CAPPC) can be obtained from Eq.(1), i.e.,

$$\theta_A^{Crit} = \theta/N_A = 360°/6 = 60° \quad (1)$$

At 150 K, the A-BPNT with chirality of (**9**, 7, 360°) is stable but not for A-BPNT (**8**, 7, 360°). Clearly, the critical value of the CAPPC (along armchair direction) is 40° at 150 K. Hence, the curvature of A-BPNT (8, 7, 360°) is higher than that of A-BPNT (6, 7, 360°) which is stable at 4.3 K. If the environmental temperature is 300 K, even the configuration of A-BPNT (15, 7, 360°) is unstable. The critical value of the CAPPC is 22.5°. It is concluded that the thermal vibration of atoms on tube is the second major reason for tube destruction.

To find the critical curvature of a Z-BPNT (e.g., Figure 1f) at finite temperature, we curl the same rectangular SLBP along zigzag direction with different curved angles. Figure 2b gives three critical states of the Z-BPNTs. At 4.3 K, the critical value of $\theta$ is 224° because the arch structure (Z-BPNT (6, 15, $\theta$)) is broken if $\theta$ is slightly higher than 224° (see the second insert snapshot in Figure 2b). Hence, the critical value of the CAPPC (along zigzag direction) can be calculated by Eq.(2), i.e.,

$$\theta_Z^{Crit} = \theta / N_Z = 224° / 15 \approx 14.9° \tag{2}$$

When the temperature is high, e.g., 150K, the critical value of $\theta$ is 265° for the arch structure of Z-BPNT (6, 31, $\theta$). Correspondingly, the critical value of the CAPPC is ~8.5°, which is far less than 14.9° at 4.3 K. At 300 K, the CAPPC is ~6.9°. Comparing the three failure models shown in Figure 2b, we also find that the damage area is nearby the center of the arch structure at low temperature, e.g., 4.3 or 150 K. However, at 300 K, the damage area (shown in the 6th snapshot in Figure 2b) is close to the lower fixed end. The reason is that the drastic thermal vibration of phosphorus atoms on the arch structure leads to a random initial damage on the structure.
From above, one concludes that the critical curvature of either an A-type or a Z-type BPNT is higher at lower temperature.

### 3.2 Failure mechanism of Z-BPNTs

From above, we also find that the critical value of the CAPPC of an A-BPNT is much higher than that of a Z-BPNT at the same temperature. Hence, the stability of a Z-BPNT is lower than an A-BPNT at the same temperature. To demonstrate the failure of a Z-BPNT due to local damage on the outer layer at finite temperature, the arch structure of Z-BPNT (6, **15**, $\theta$) is chosen and the bond lengths and bond angles on the outer layer nearby the initial fracture area (see Figure 3) are calculated because the bonds are under tension during deformation.

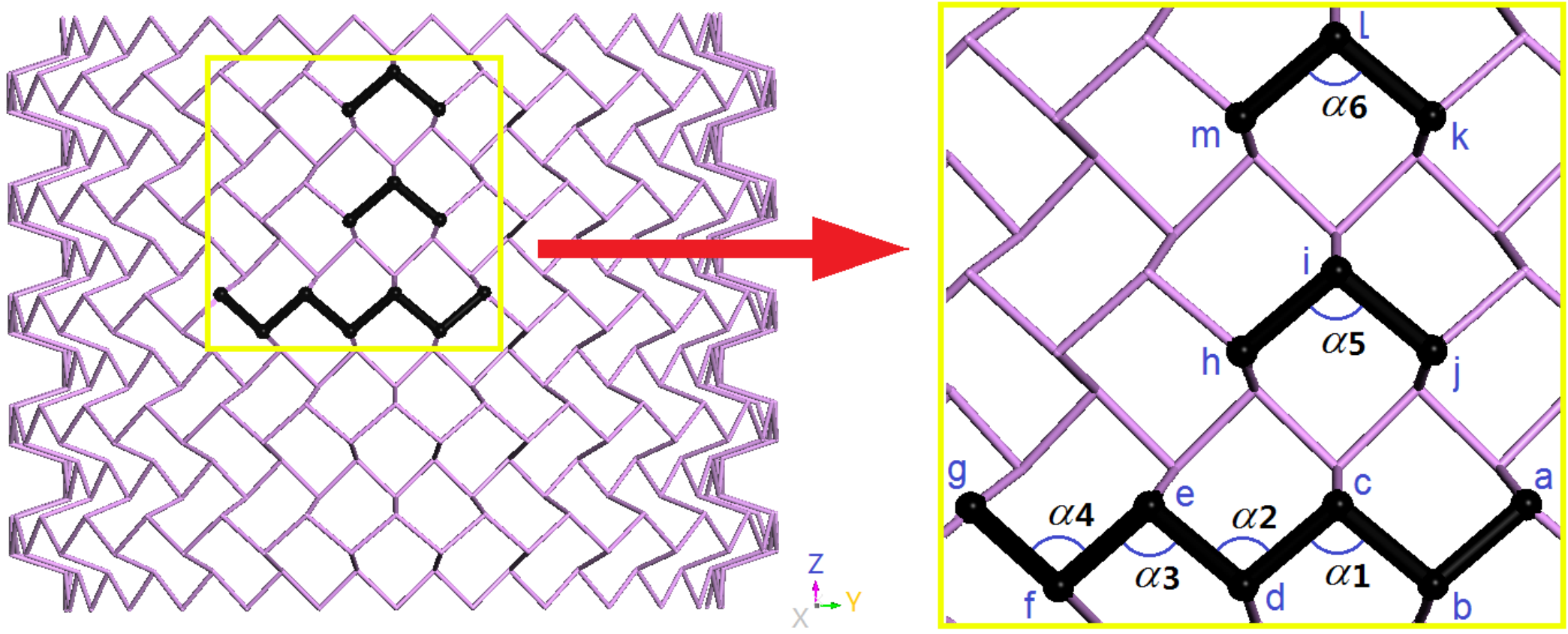

**Figure 3.** Initial configuration of Z-BPNT (6, 15, $\theta$) at 4.3 K and some typical bonds and bond angles on the outer layer of the tube. Along axial/z-direction, there are three angles i.e., $\alpha 1$, $\alpha 5$ and $\alpha 6$, and the six lengths of bonds b-c, c-d, i-j, i-h, l-k and l-m. Along circular-direction, the four angles, i.e., a1, $\alpha 2$, $\alpha 3$ and $\alpha 4$, and six lengths of bonds a-b, b-c, c-d, d-e, e-f and f-g, are calculated and compared.

Figure 2b implies that the Z-BPNT (6, 15, $\theta$) at 4.3 K is in the critical stable state when $\theta$=224°. It is known that the temperature effect, i.e., thermal vibration of atoms, is negligible for the arch structure at such low temperature. Hence, the failure of the arch structure is mainly caused by tensile deformation if the value of $\theta$ increases slightly, e.g., $\theta$=225°. To reveal the initial damage source of the arch structure, the variations of the bond lengths and bond angles nearby the failure area (Figure 3) are calculated. Figure 4 shows the histories of the 10 bond lengths and 4 angles on the outer layer of the arch structure at either stable state ($\theta$=224°) or failure state ($\theta$=225°).

When $\theta$=224°, all of the 10 bonds have very small variation and are less than 0.25 nm (Figure 4a and c), which is less than the maximal bond length (0.279 nm). Among the six bond angles, $\alpha 1$, which is at the center of the outer layer of the arch structure, has the highest variation. Meanwhile, the variation of the bond angle is lower when the distance between the center atom and atom c (the center atom of $\alpha 1$) is higher (Figure 4b and d). When $\theta$=225°, $L_{bc}$, the length of bond b-c (between atoms b and c; similar to other bonds), jumps up suddenly after 2750 fs. It reaches the maximum of bond length, i.e., 0.279 nm, at 2820 fs. After then, bond b-c breaks and the local fracture keeps developing. Simultaneously, the lengths of the rest bonds, e.g., $L_{cd}$, $L_{de}$, decrease. The reason is that the area nearby broken bond b-c becomes a new free boundary, and the initial tensile deformation of the rest bonds nearby the new boundary drops (Figure 4a and c). Comparing the variation of $\alpha 1$, $\alpha 5$ and $\alpha 6$ (Figure 4b), we find that the bond angles along axial-direction vary similarly to those in the structure with $\theta$=224°. However, along the circular-direction, the values of $\alpha 2$, $\alpha 3$ and $\alpha 4$ drop sharply when bond b-c breaks. It also attributes to appearing of the new boundary nearby the broken bond b-c.

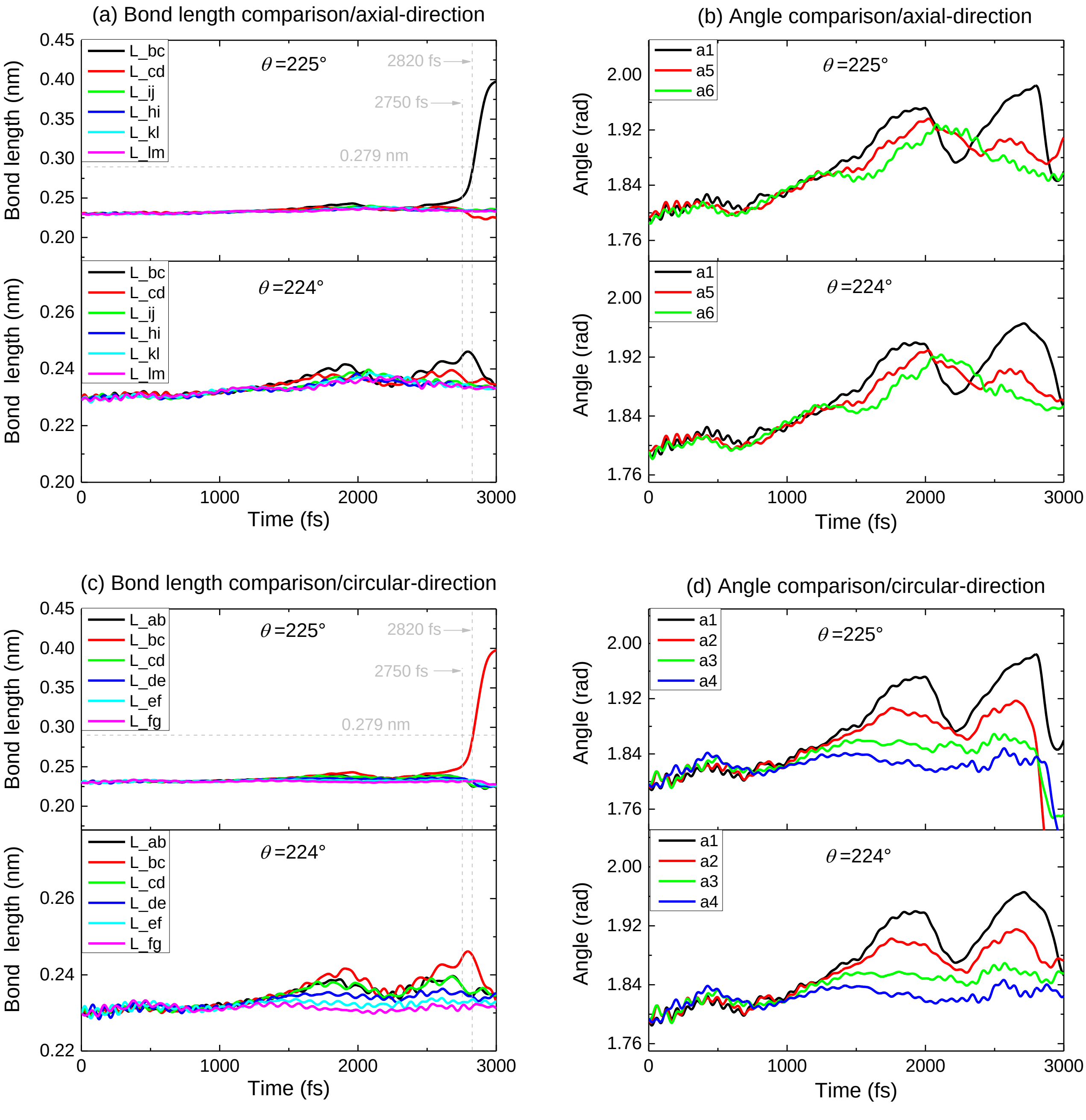

**Figure 4.** Histories of bond angles and bond lengths in the Z-BPNTs of (6, 15, $\theta$) at 4.3 K with different curved

angle, i.e., $\theta$=224° and $\theta$=225°. L_xy=$L_{xy}$ (x, y=a, b, c, d, e, f, g, h, i, j, k, l, m). (a) Bond lengths comparison along axial direction; (b) bond angles comparison along axial direction; (c) bond lengths comparison along circular direction; (b) bond angles comparison along circular direction.

In the SW potential for BP, the distance between two non-bonded atoms, e.g., $L_{bd}$, also has maximum, i.e., 0.389 nm [17]. In the above analysis, we only find the difference between the variation histories of bond lengths and bond angles nearby the center of the out layer of the arch structure. We only find the critical curvature of the arch structure. The reason for the fracture of bond b-c is not clear yet. In Figure 5, we provide the histories of $L_{bc}$ and $\alpha$1 during [2500, 3000] fs. The curves indicate that $\alpha$1 approaches maximum (1.9657 rad at Peak 1 in the upper layer of Figure 5) at 2710 fs when $\theta$=224°. Meanwhile, $L_{bc}$=0.2422 nm, $L_{cd}$=0.2389 nm, and $L_{bd}$, i.e., the distance between atoms b and d, can be obtained by the formulation

$$L_{bd} = \sqrt{L_{bc}^2 + L_{cd}^2 - 2L_{bc}L_{cd}\cos(\alpha 1)} \ . \quad (3)$$

$L_{bd}$ is 0.4003 nm, which is greater than its maximum value (0.389 nm). Fortunately, the value of $\alpha$1 drops from then on and the bond lengths, i.e., $L_{bc}$ and $L_{cd}$, are much lower than their maximal value (0.279 nm). Hence, the structure keeps undamaged.

For the case of $\theta$=225°, at 2710 fs, $\alpha$1=1.976 rad, $L_{bc}$=0.2469 nm (<0.279 nm), $L_{cd}$=0.2373 nm (<0.279 nm) and $L_{bd}$=0.4043 nm >0.389 nm. However, $\alpha$1 keeps increasing before 2798 fs (at Peak 2 in the upper layer of Figure 5). Especially, $L_{bc}$ increases continuously before 3000 fs. At 2820 fs, $L_{bc}$ reaches its maximal value and bond b-c is broken. It implies that the local stability of the arch structure loses due to $L_{bd}$ >0.379 nm when $\theta$ is over 224°, rather than due to thermal vibration of atoms.

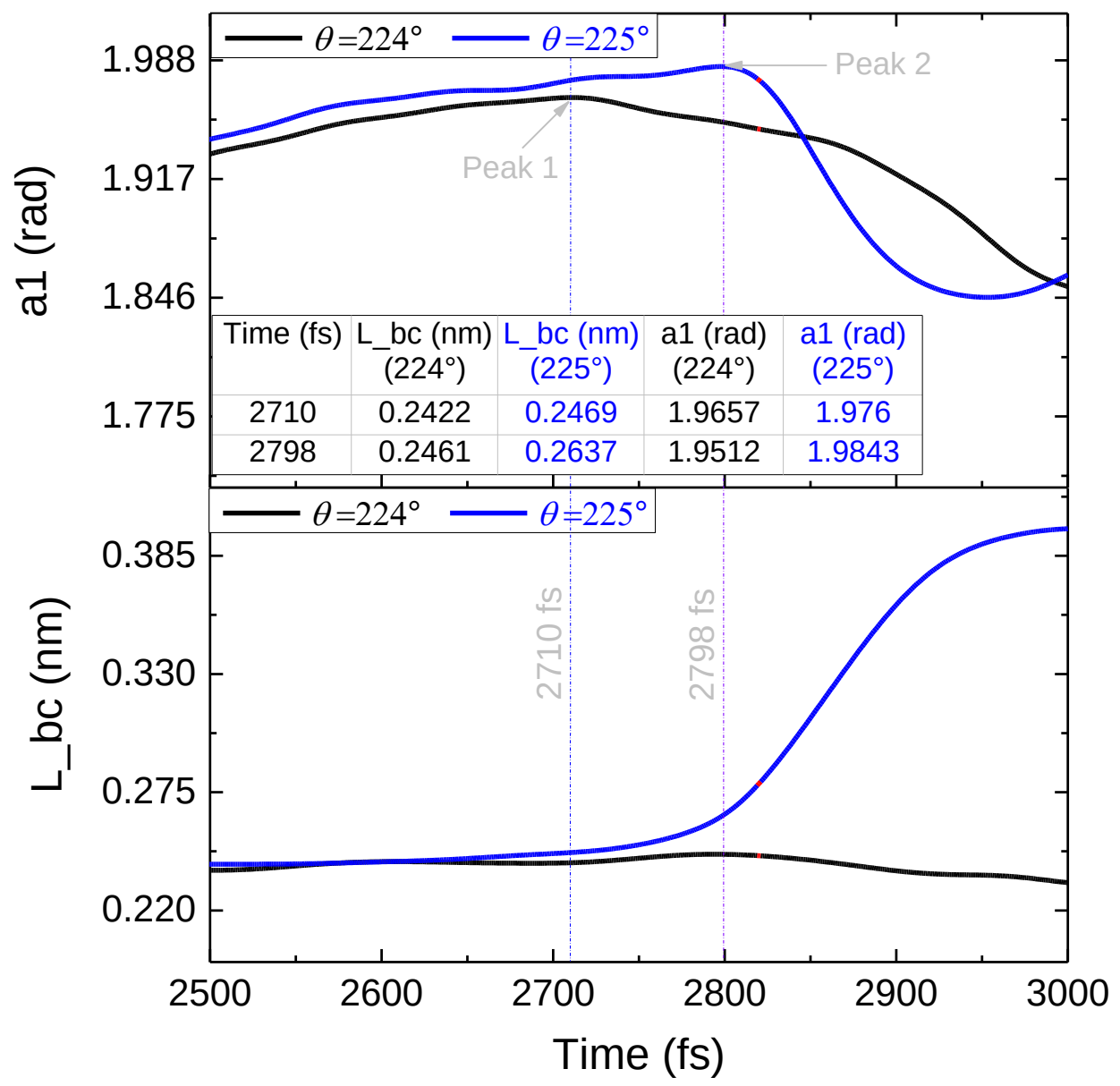

**Figure 5.** Histories of $\alpha$1 and L_bc (=$L_{bc}$) of the structure under critical states at 4.3 K.

### 3.3 Temperature effect on failure of BPNTs

Figure 2 indicates that the critical CAPPC of the same BPNT is lower at higher temperature. In the above discussion, the effect of thermal vibration of atoms on the failure of BPNT at 4.3 K is neglected. Now, we want to know the relationship between the critical CAPPC of a Z-BPNT and the environmental temperature. With the understanding of this relationship, one can design a stable BPNT in a nano device. The relationship is analyzed via the following steps.

(1) Pure thermal deformation of a P-P bond

From the three fractured structures shown in Figure 2b, we know that the initial damage of the tube is random as the temperature is high. It implies that the first breakage of bond is caused by the highest thermal vibration of the atoms on the bond. Hence, the maximal thermal vibration induced elongation of P-P bonds on SLBP at finite temperature should be obtained. For finding the maximal elongation at specified temperature, we fix one pucker side of a rectangular BP sheet with a size of 10.4 nm-by-10.6 nm, and in statistics the bonds are more than 2.0 nm far away from the four sides of the sheet during 200 ps of relaxation at a NVT ensemble.

Figure 6 indicates that the thermal vibration induced bond elongation increases with the increasing of temperature. The engineering strain of a P-P bond is over 10% when the environmental temperature is over 200 K. Hence, the serious variation of bond length due to temperature influences the mechanical stability of BPNTs.

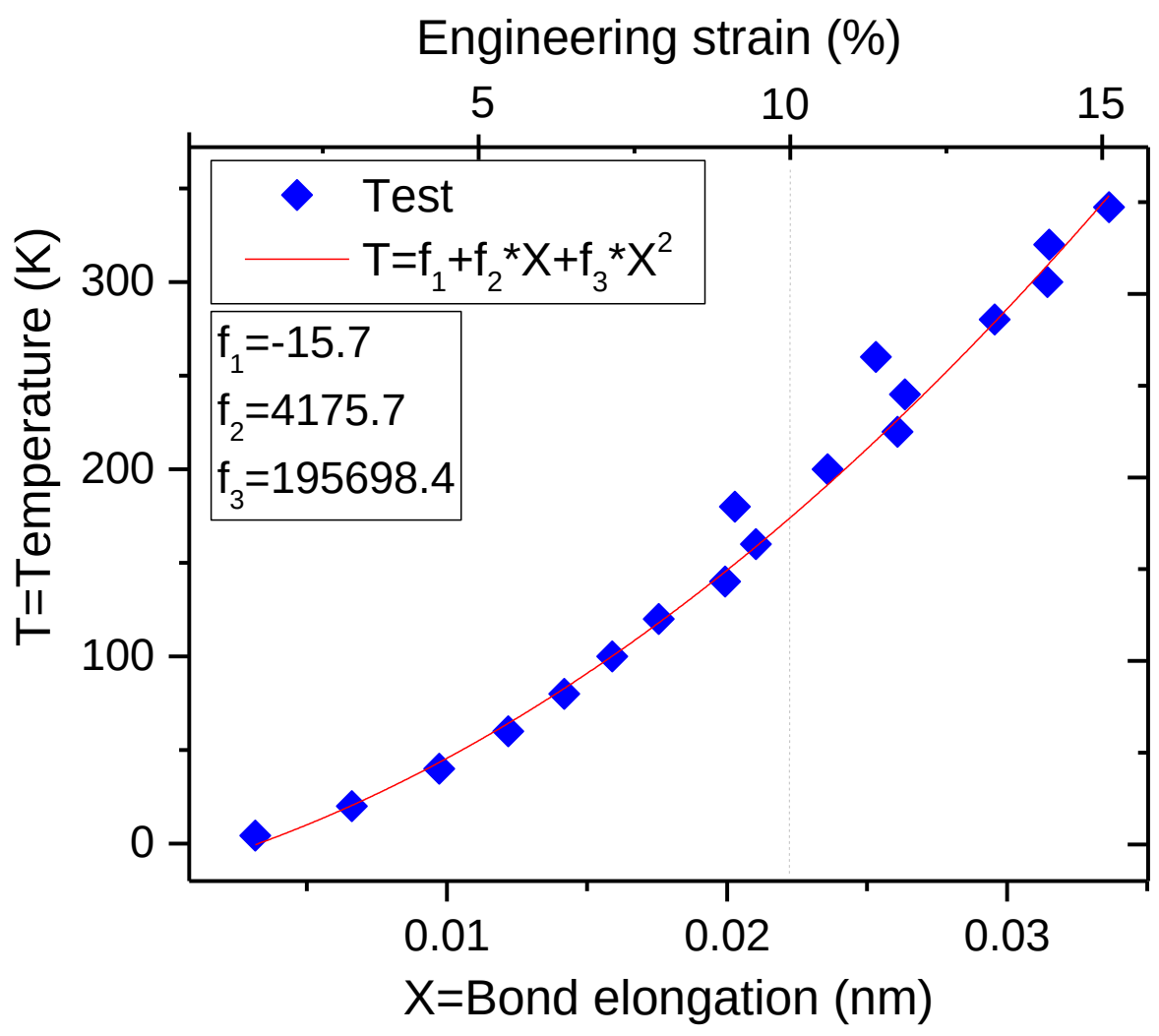

**Figure 6.** Curve of temperature v.s. maximal bond elongation due to thermal vibration of atoms on SLBP. Temperature varies in [4.3, 340] K.

From the fitting function listed in Figure 6 or Eq.(4), one can obtain the amplitude of thermal vibration of P-P bond at finite temperature (T).

$$\Delta\mathrm{L}_{45}^{\mathrm{Temp}} = \Delta\mathrm{L}_{46}^{\mathrm{Temp}} = X = \left(-f_2 + \sqrt{f_2^2 - 4f_3(f_1 - T)}\right)/(2f_3) \quad (4)$$

where $f_1$, $f_2$ and $f_3$ are fitting parameters given in Figure 6.

(2) Pure tensile deformation

The curvature induced tensile deformation of the outer layer, for example, of a BPNT (Figure 7), can be calculated. It is known that $\widehat{\mathrm{bd}}$, the arc length between atoms b and d on a Z-BPNT (Figure 7c), is higher than their distance ($L_{56}$) on the free SLBP sheet at 0 K. The difference can be calculated from Eq.(5), i.e.,

$$\Delta\mathrm{L}_{56}^{\mathrm{Tensile}} = \Delta\mathrm{L}_{\mathrm{bd}}^{\mathrm{Tensile}} = \widehat{\mathrm{bd}} - \mathrm{L}_{56} = \frac{\mathrm{L}_{14} * \cos\beta * \theta_Z^{Crit}}{2}\ , \quad (5)$$

where $L_{14}$, $L_{56}$ and $\beta$ are introduced in the caption of Figure 1. As the length variation of $L_{56}$ is mainly due to the variation of angle $\alpha 1$, it can be calculated from

$$\Delta\alpha 1 \approx \sin^{-1}\frac{\Delta L_{56}^{\mathrm{Tensile}} * \cos(0.5\alpha)}{L_{45}} = \sin^{-1}\frac{\cos(0.5\alpha) * L_{14} * \cos\beta * \theta_Z^{Crit}}{2L_{45}}\ , \tag{6}$$

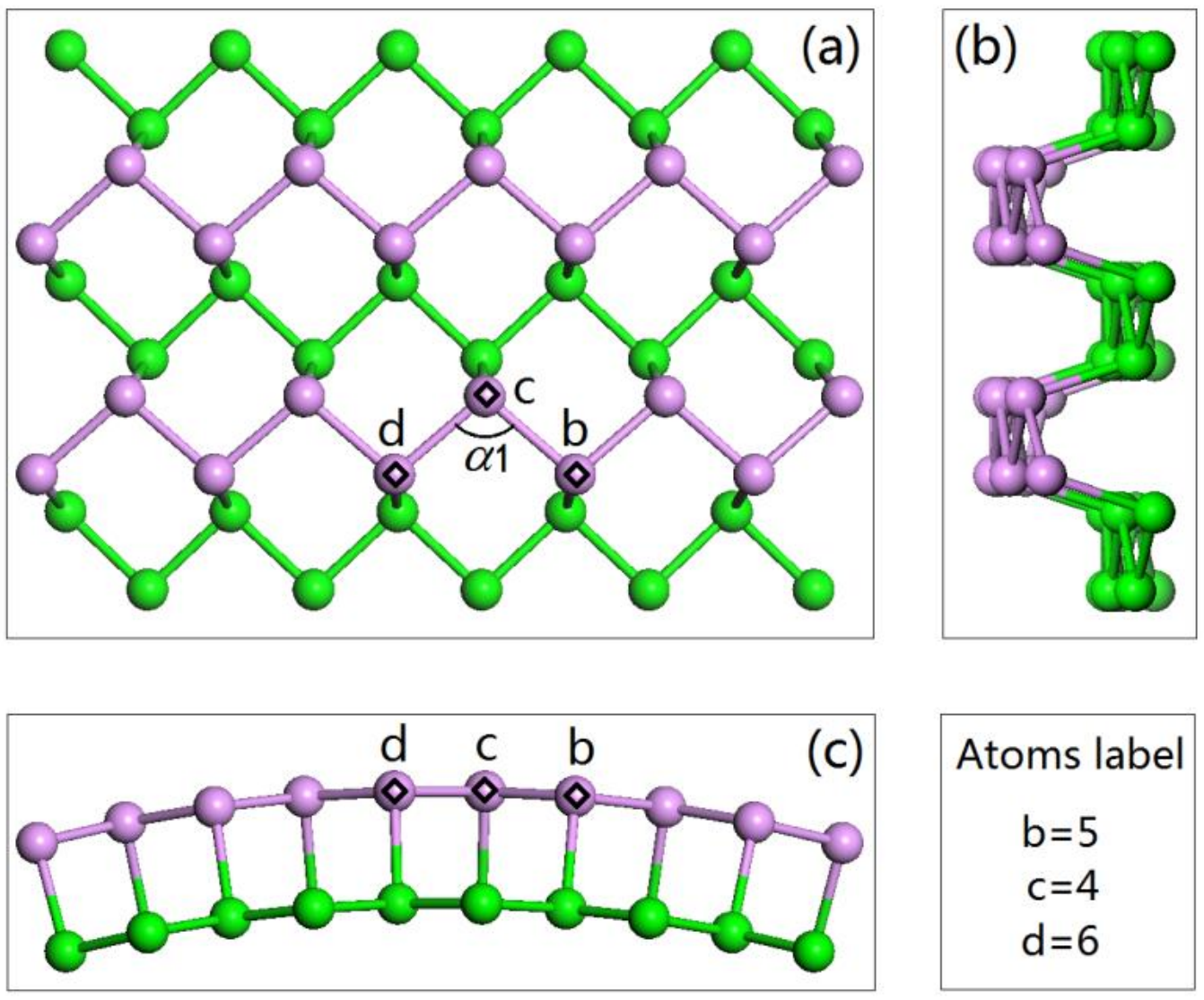

**Figure 7.** Three-view of a segment of a Z-BPNT. The three (purple) atoms, i.e., b( or 5), c(or 4) and d(or 6), and bond angle $\alpha 1$ on the outer layer of tube. Letters of b, c and d represent the atoms on the BPNT, numbers of 4, 5 and 6 represents the atoms on SLBP.

(3) Conditions for a stable BPNT

To maintain the stability of a BPNT, the three conditions in Eq. (7) should be satisfied, simultaneously,

$$\begin{cases} L_{45*} = L_{45} + \Delta L_{45}^{\mathrm{Tensile}} + \Delta L_{45}^{\mathrm{Temp}} < 0.279\ \mathrm{nm}, \\ L_{46*} = L_{46} + \Delta L_{46}^{\mathrm{Tensile}} + \Delta L_{46}^{\mathrm{Temp}} < 0.279\ \mathrm{nm}, \\ L_{56*} = \sqrt{L_{45*}^2 + L_{46*}^2 - 2L_{45*}L_{46*}\cos(\alpha + \Delta\alpha 1)} < 0.389\ \mathrm{nm}. \end{cases} \tag{7}$$

The first two inequations imply that no bond on the BPNT is broken at finite temperature. The third one demonstrates that the variation of $\alpha 1$ should be low when the two bonds 4-5 and 4-6 are under tension.

In Eq.(7), $L_{45}$ and $L_{46}$ are constants. Hence, one can find that a BPNT at a higher temperature will bear lower tensile deformation. That is the reason that the critical value of CAPPC of a BPNT decreases with the increasing of environmental temperature. From Eq.(7) one can obtain the coupling effect of temperature effect and curvature of a stable BPNT.

(4) Robustness design of a stable BPNT

For a BPNT used in a practical application, the tube should be far away from its critical stable state. The above results demonstrate that both of the environmental temperature and curvature have significant influence on the stable state. Therefore, the robustness of a stable BPNT can be expressed by the capacity of the tube when temperature becomes higher as it is working. For example, if a BPNT is stable at temperature T, it still undamaged at T+ΔT. The positive increment, ΔT, indicates the robustness of the tube. In comparing, if the environmental temperature is well controlled, the number of periodic cells on the cross section (i.e., along the circular direction) of the tube should be greater than the minimal number of periodic cells of the tube at the temperature (Figure 8). The difference between the two numbers also demonstrates the robustness of the BPNT.

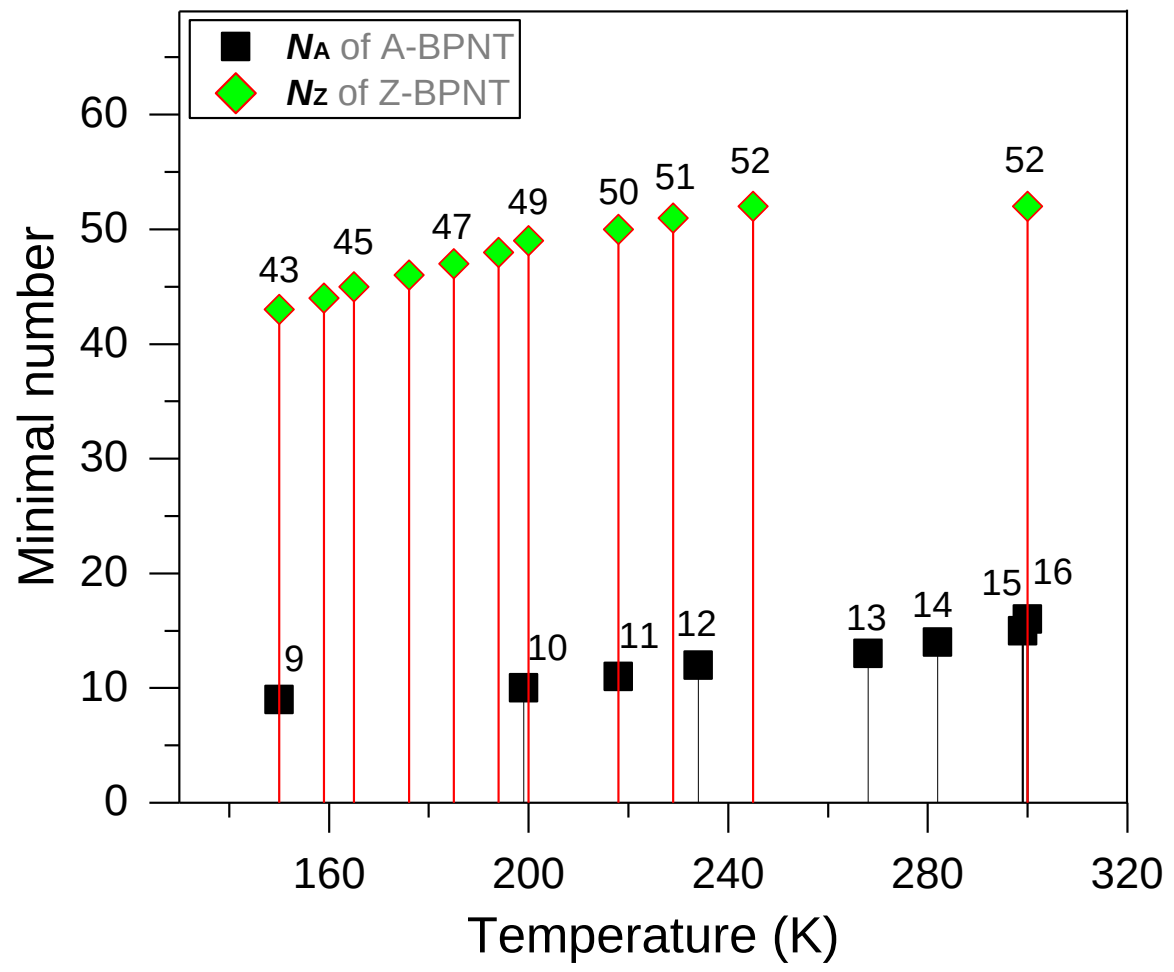

**Figure 8.** The minimal number of periodic cells in the circular direction of a BPNT of either A-type ($N_A$, --, 360°) or Z-type (--, $N_Z$, 360°) at finite temperature between 150 and 300 K. The critical temperature of $N_A$ or $N_Z$ is found by bisection method between 150 and 300 K, i.e., the BPNT is at the critical stable state at the temperature. The tube is relaxed for 800 ps (time step is 0.1 fs) at a canonical NVT ensemble at each temperature. The edge of Z-BPNT is unstable. To avoid initial damage of the edge at a high temperature after energy minimization, the temperature of the tube increases linearly from 4.3 K to the value during bisection within 500 ps.

# 4 Conclusions

To find the effect of temperature on the failure of a BPNT at canonical NVT ensemble, a series of MD simulations are presented. In simulation, two types of BPNTs are obtained by curling the rectangular SLBP sheet along different directions. From the results, some concluding remarks are drawn.

(1) Bond elongation due to thermal vibration increases with the increasing of temperature. The critical curvature of either an A-type or a Z-type BPNT is higher at lower temperature;

(2) The failure of a Z-BPNT at a specified temperature is mainly due to the buckling of angle $\alpha$1, which locates in the middle part of the tube. For an A-BPNT, the failure of tube is caused by both of the breakage of P-P bonds on the outer layer of tube and the strong attraction of the atoms on the inner layer;

(3) For a stable BPNT at finite temperature, all the bond lengths should be less than 0.279 nm, and the smallest distance between non-bonded atoms should be less than its maximal value, i.e., 0.389 nm, simultaneously.

From the application point of view, a BPNT should have a lower curvature than its critical value at the same temperature. A BPNT with higher radius or at lower temperature will have higher robustness in acting as a component in a nano device.

# Acknowledgements

The authors are grateful for the financial support from the National Natural-Science-Foundation of China (Grant

Nos. 11502217, 11372100) and the Australian Research Council (Grant No. DP140103137).